\documentclass[conference]{IEEEtran}
\pdfoutput=1
\usepackage{graphicx}
\usepackage{xspace}
\usepackage{cite}
\usepackage{subcaption}
\usepackage{footnote}
\usepackage{mathtools}
\usepackage{url}
\usepackage{chngcntr}
\usepackage{listings, color, caption}
\usepackage{threeparttable}
\usepackage{setspace}
\usepackage{multirow}
\usepackage{multicol}
\usepackage{pifont}
\usepackage{soul}
\usepackage{boxedminipage}
\usepackage{booktabs}
\usepackage{multirow}
\usepackage{hyperref}
\usepackage{stfloats}

\usepackage{algorithm,algorithmicx}
\usepackage[noend]{algpseudocode}
\usepackage{indentfirst}
\usepackage{comment}
\graphicspath{ {images/} }
\usepackage{amsfonts}
\usepackage{mathrsfs}
\usepackage{amssymb}
\usepackage{amsthm}

\newcommand{\RN}[1]{%
  \textup{\uppercase\expandafter{\romannumeral#1}}%
}

\algrenewcommand\algorithmicindent{0.7em}%

\title{Background-aware Multi-source Fusion Financial Trend Forecasting Mechanism}

\author{
 \IEEEauthorblockN{Fengting Mo\IEEEauthorrefmark{2}, Shanshan Yan\IEEEauthorrefmark{2},  Yinhao Xiao \IEEEauthorrefmark{1}\IEEEauthorrefmark{2}}

 \IEEEauthorblockA{\IEEEauthorrefmark{2}School of Information Science, Guangdong University of Finance and Economics, Guangzhou, China.}

 \IEEEauthorblockA{\IEEEauthorrefmark{1}Corresponding Author}

}

\begin{document}

\maketitle
\begin{abstract}
With the rapid development of economic globalization and the emergence of digital economics, the global economy has been propelled forward. Stock prices, as an economic indicator, reflect changes in economic development and market conditions. However, stock prices are characterized by high risk, high noise, and frequent fluctuations, making accurate prediction of stock price movements challenging. Enhancing the accuracy of stock price prediction has become a focal point of attention for many scholars and investors.\par
Traditional stock price prediction models often only consider time-series data and are limited by the mechanisms of the models themselves. Some deep learning models have high computational costs, depend on a large amount of high-quality data, and have poor interpretations, making it difficult to intuitively understand the driving factors behind the predictions. Some studies have used deep learning models to extract text features and combine them with price data to make joint predictions, but there are issues with dealing with information noise, accurate extraction of text sentiment, and how to efficiently fuse text and numerical data. In addition, the models are highly sensitive to the quality and real-time availability of textual data, which may limit the generality and stability of the models.

To address these issues in this paper, we propose a background-aware multi-source fusion financial trend forecasting mechanism. The system leverages a large language model to extract key information from policy and stock review texts, utilizing the MacBERT model to generate feature vectors. These vectors are then integrated with stock price data to form comprehensive feature representations. These integrated features are input into a neural network comprising various deep learning architectures. By integrating multiple data sources, the system offers a holistic view of market dynamics. It harnesses the comprehensive analytical and interpretative capabilities of large language models, retaining deep semantic and sentiment information from policy texts to provide richer input features for stock trend prediction. This integration effectively enhances the accuracy and interpretability of stock price predictions.

Additionally, we compare the accuracy of six models (LSTM, BiLSTM, MogrifierLSTM, GRU, ST-LSTM, SwinLSTM). The results demonstrate that our system achieves generally better accuracy in predicting stock movements, attributed to the incorporation of large language model processing, policy information, and other influential features. The multi-dimensional and multi-level fusion forecasting mechanism not only improves forecast accuracy but also enhances the model's generalization ability and interpretability. These advancements represent significant strides toward more efficient and intelligent financial market forecasting.

\end{abstract}

\section{Introduction}
\label{sec:introduction}

With the economic prosperity and the rapid emergence of the big data industry, the stock market, as an important part of the national economic landscape, has been receiving more and more attention from the public. Stock price forecasting has been a hot research topic in the financial field. However, stock prices not only exhibit high volatility and noise, but are also affected by factors such as national exchange rates, fiscal policies, market sentiment and macroeconomic conditions. As a result, analysing and forecasting stock prices poses significant challenges, making them a focus of research and attention for investors. In academia, research on stock price forecasting has been studied continuously for many years. In earlier studies, some of the main research methods used in academia to predict stock prices include statistical models, machine learning methods, deep learning methods, and sentiment analysis methods.\par

Traditional statistical models predict new signals by linearly combining historical signals and independent noise terms, including time series analysis, regression analysis, such as autoregressive integrated moving average model (ARIMA)~\cite{box2015time} and generalised autoregressive conditional hetero-skedasticity model (GARCH)~\cite{francq2019garch} and stochastic volatility model among others. However these traditional models underperform in the face of complex, non-linear and high-frequency data from the stock market. In addition, traditional statistical methods are sensitive to data variations and outliers, making them susceptible to the influence of noisy data. This can lead to significant prediction errors when dealing with real-time, dynamic stock price data.\par
With the rise of cutting-edge technologies such as artificial intelligence and big data, machine learning methods and Deep learning mothods have been widely applied in stock price prediction. Machine learning methods offer advantages such as handling large volumes of data, capturing complex nonlinear relationships, and automatically learning feature representations. Examples include logistic regression models (LR)~\cite{wu2021cash}, support vector machines (SVM)~\cite{gang2019safety}, gradient boosting decision trees (GBDT)~\cite{zhou2019cascading}, and neural networks. 
Deep learning models possess powerful feature learning and pattern recognition capabilities, enabling them to automatically learn complex feature representations from large-scale data and achieve more accurate predictions. Compared to machine learning methods, deep learning models offer higher flexibility and scalability, allowing them to adapt to different prediction tasks and data types by increasing network layers and adjusting network structures. They are better suited to handle complex nonlinear relationships, making them more appropriate for stock trend prediction.
The application of deep learning in stock price prediction includes models such as Recurrent Neural Networks (RNN)~\cite{elman1990finding}, Long Short-Term Memory networks (LSTM)~\cite{hochreiter1997long}, Convolutional Neural Networks (CNN)~\cite{lecun1998gradient}, Deep Neural Networks (DNN) and Transformer~\cite{vaswani2017attention}. These models can learn feature representations from historical stock price data and predict future stock price movements through time series analysis and pattern recognition. RNN is a neural network structure capable of processing sequential data, suitable for modeling and predicting stock time series data. It captures temporal dependencies in the data and can flexibly handle variable-length input sequences. LSTM is an improved RNN structure that introduces gated units to more effectively capture and retain long-term dependencies. In stock price prediction, LSTM can efficiently handle long-term time series data and has strong representation capabilities. CNN, which has achieved great success in image processing, can also be applied to stock prediction. Through convolution operations, it can extract local features from time series data, helping to identify patterns and trends in stock price changes.DNN is a multi-layer neural network structure that can learn complex feature representations through multiple layers of nonlinear transformations. In stock price prediction, DNN can learn higher-order feature representations from historical stock price data, improving prediction accuracy.The core component of Transformer is the self-attention mechanism, which allows the model to consider all elements in the entire sequence when processing the input sequence. This allows the model to capture long distance dependencies between different positions in the sequence.Moreover, Transformer does not rely on sequential processing and can process the entire sequence in parallel. It is able to enhance the prediction ability by utilising its powerful feature extraction capabilities and efficient processing.

Yet while some deep learning models have made breakthroughs in predictive capabilities, they suffer from high computational costs, high risk of overfitting, and dependence on large amounts of high-quality data. There are also issues such as poor model interpretability and difficulty in intuitively understanding the drivers behind predictions.\par

In the academic field, sentiment analysis methods for stock price prediction aim to forecast stock market trends by analyzing sentiment information in text data combined with stock data. These methods typically utilize unstructured data from social media, news reports, financial commentaries, and other sources. Through text mining and sentiment analysis techniques, they extract sentiment signals and perform correlation analysis with stock price data. However, the accuracy of sentiment analysis can be affected by factors such as the quality of the text data, misinterpretation of context, and the difficulty of capturing the complex relationship between market sentiment and stock prices. \par

Some studies have used deep learning models to extract text features and combine them with price data to make joint predictions, but there are problems in dealing with information noise, accurate extraction of text sentiment, and how to efficiently fuse text and numerical data. In addition, the models are highly sensitive to the quality and real-time availability of text data, which may limit the generality and stability of the models.\par

To address these issues, this paper proposes a background-aware multi-source fusion financial trend forecasting mechanism. The mechanism utilises a large language model to extract key information from policies and stock reviews, and uses the MacBERT model to form feature vectors, which are then combined with stock price data for prediction. By integrating multiple data sources, the mechanism provides a comprehensive understanding of market dynamics. It fully utilizes the comprehensive analysis and interpretation capabilities of large language models, retaining the deep semantic and sentiment information of the text, thus offering richer input features for stock trend prediction. This mechanism effectively enhances the accuracy and interpretability of the stock price prediction system. The mechanism has achieved significant predictive performance in experiments, providing new insights and methods for stock market prediction and decision-making.

Our contributions are fourfold:
\begin{itemize}
	\item  We utilize the comprehensive analytical understanding and information extraction capabilities of large language models to summarize policy texts. This approach distills key information from policy texts and effectively enhancing the prediction mechanism's sensitivity to policy impacts. 

 \item We uses the pre-trained Chinese language model MacBERT to extract the feature vectors of policy summaries and stock comment summaries and fuse them with stock data. This fusion helps to enrich the data dimensions, fully consider the deep semantic and sentiment information in different texts, and address the shortcomings of traditional sentiment classification methods that cannot understand the text content and its potential impact.
	\item To account for the influence of past stock movements on future trends, the model incorporates a Prior Effect module.  This module enables the mechanism to consider historical stock movements when making predictions, thus enhancing the model’s ability to learn from sequential data.This effectively improves the accuracy and interpretability of the stock price prediction system.

	\item We propose a background-aware multi-source fusion financial trend forecasting mechanism. The mechanism integrates the advantages of natural language processing and time series analysis, forming a systematic and integrated prediction mechanism that provides investors with more reliable and comprehensive predictive information.

\end{itemize}

\textbf{Paper Organization.} 
The rest of the paper is organized as follows. Section~\ref{sec:background} presents the recently advanced background knowledge of our approach. Section~\ref{sec:approach} details the approach components of the mechanism. Section~\ref{sec:eval} reports our experimental results on our system. Section~\ref{sec:related} outlines the most related work.  Section~\ref{sec:conclusion} concludes the paper with a future research discussion.

\section{Background}
\label{sec:background} 
In this section, we mainly demonstrate the background knowledge of some recently advanced technologies.

\subsection{Pre-trained Models}
Pre-trained Models (PTM)~\cite{han2021pre} refer to neural network models that have been previously trained on large-scale datasets. These models typically learn features on general tasks that can be transferred to other specific tasks. The advantage of pre-trained models is that they can utilize the statistical information from large-scale data to learn universal language representations, without the need for expensive data labeling and model training for specific tasks. Once pre-training is completed, these models can be easily fine-tuned or transferred to specific tasks to meet the requirements of particular domains or tasks.

In recent years, due to the development of deep learning and large-scale data, pre-trained models have achieved great success in the field of Natural Language Processing (NLP). The emergence of Transformers has particularly revolutionized the landscape of pre-trained models. Transformer models achieve better contextual understanding through self-attention mechanisms, enabling them to better capture the global semantic information of sentences and documents. BERT~\cite{devlin2018bert}, GPT~\cite{radford2018improving}, and other Transformer-based models have started to be widely applied in NLP tasks. These large-scale PTMs can capture multiple disambiguations, lexical and syntactic structures, and factual knowledge from texts. By fine-tuning these large-scale PTMs on large samples, their rich linguistic knowledge brings superior performance to downstream NLP tasks.
\subsection{Large language model (LLM)}
A Large Language Model (LLM) is a deep learning model capable of understanding and generating natural language text. It is usually based on a neural network architecture with a large number of parameters and a rich language representation.

\par A significant advantage of LLMs is their powerful generative capabilities. By learning large amounts of textual data, these models are able to generate high-quality, coherent text. Multi-tasking capability is another major advantage of big language models. While traditional NLP models usually need to be trained individually for each task, LLMs are able to achieve excellent performance on multiple tasks through pre-training and fine-tuning. This multi-tasking capability not only improves the efficiency of model application, but also reduces development cost and time. Typical tasks include text generation, translation, Q\&A systems, summary generation, sentiment analysis and text classification. Big Language Models can be quickly adapted to new tasks through a simple fine-tuning process, demonstrating flexibility and adaptability.

\par In terms of language understanding, LLMs show deep analysis and comprehension capabilities. They are able to capture complex patterns and long-distance dependencies in a language and provide comprehensive understanding and processing of contextual information. This ability allows LLMs to excel in tasks such as reading comprehension, information extraction, and dialogue systems. For example, the BERT model is able to understand the relationship between each word in a sentence and other words through a bidirectional encoder structure, thus providing more accurate semantic understanding.
\par Meanwhile, LLM is scalable, and the performance of LLM can be significantly improved by increasing the number of parameters and the size of training data of the model. GPT-3, for example, uses 1.5 trillion parameters, which is 10 times more than GPT-3.5, and is more creative and capable of handling more nuanced instructions. With the continuous growth of computational resources and data size, the potential of LLMs will be further released to support more complex and advanced language tasks.
\subsection{MacBERT}
MacBERT~\cite{cui2020revisiting} is an improved version of the BERT-based model designed to improve the performance of Chinese language processing tasks. It replaces the original MLM task as MLM as Correction (Mac) task by masking words using words similar to them, reducing the gap between the pre-training and fine-tuning phases. This masking strategy has been shown to be very effective in downstream tasks.MacBERT is designed to provide a powerful and effective pre-trained language model to advance the field of Chinese natural language processing research.\par MacBERT employs larger and more diverse Chinese text data for pre-training to improve the model's comprehension of Chinese language. Different from traditional BERT models, MacBERT innovates in masking strategy. Instead of using the traditional [MASK] notation to mask the target word, it uses words similar to the target word for replacement. This similar word substitution strategy helps to better retain semantic information and improves the performance of the model in fine-tuning tasks. In addition, MacBERT employs whole-word masking and N-gram masking strategies to enhance the model's comprehension of long texts and improve the ability to capture contextual information. For tasks similar to the NSP task, the Sentence Order Prediction (SOP) task is performed. These designs enable MacBERT to perform well in handling various Chinese natural language processing tasks.

\subsection{Self-Attention Mechanism}
The Self-Attention Mechanism~\cite{vaswani2017attention} is an important technique widely used in sequence data processing. Originally introduced and extensively utilized in the Transformer model, the Self-Attention Mechanism has found widespread applications, particularly in the field of natural language processing, including tasks such as machine translation, text classification, and language modeling. Its emergence has profoundly altered the way sequence modeling is conducted, providing models with enhanced capabilities to better understand sequential data.

\par In the self-attention mechanism, given an input sequence, three types of vectors need to be generated for each element: Query Vectors, Key Vectors, and Value Vectors. For each Query Vector, a set of attention weights is obtained by calculating its similarity with all Key Vectors. During this calculation, the softmax function is introduced to ensure the normalization of the attention weights. The attention weights for each Query Vector are then multiplied by the corresponding Value Vectors and summed to produce the output representation of the Query Vector. This output representation can be viewed as a weighted summary of the input sequence, where the weights are dynamically generated by the attention mechanism.

\par The advantage of the self-attention mechanism is that it can consider information from all other positions at each position, thereby capturing global dependencies. It can dynamically assign weights to each element, adjusting the weight distribution based on contextual information, thus better understanding the complex relationships within the sequence. It is flexible, applicable to sequences of different lengths, and not restricted by a fixed window size, providing strong flexibility and versatility. Its computational process can also be highly parallelized, effectively leveraging the parallel computing capabilities of modern computational devices, accelerating the training and inference processes of the model.

\section{Approach}
\label{sec:approach}

\begin{figure*}[htbp]
	\includegraphics[width=190mm]{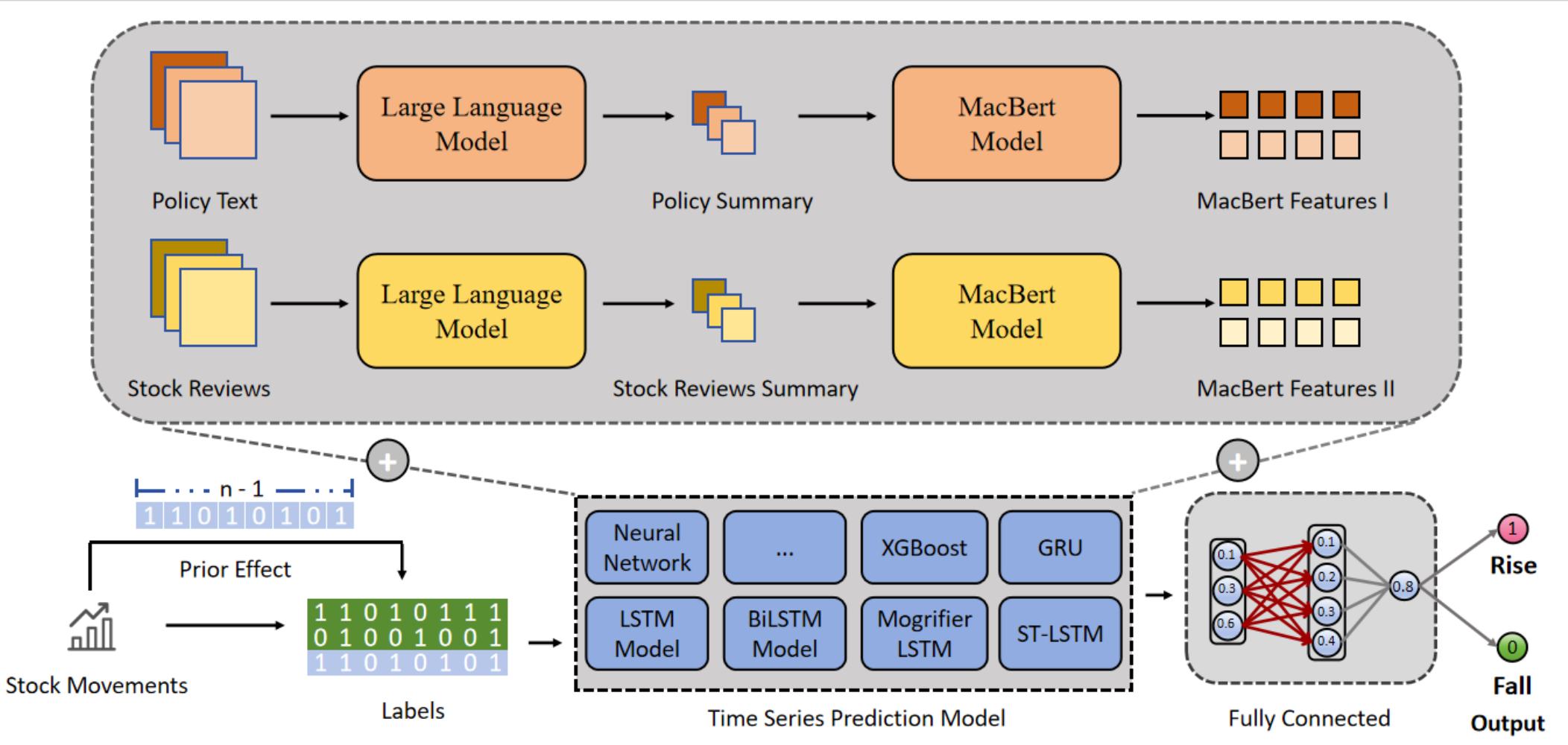}
	\centering
	\caption{Model Architecture}
	\label{fig:model}
\end{figure*}

In this section, we detail the design of the Background-aware multi-source fusion financial trend forecasting mechanism, hereinafter referred to as MOF. The overall structure of MOF is shown in Fig.~\ref{fig:model}. 
The workflow of MOF can be described as a four-step process. First, MOF takes preprocessed stock price movement data as input, obtaining information such as stock date (Date) and price change (Chg), and represents it as binary labels of 0 and 1. Second, MOF uses a sliding window to capture the price movements of the preceding n-1 days relative to the target day, representing them as binary labels, and combines this with the subsequent price movement data to be predicted. Third, MOF further extracts features from textual information such as policy information and stock comments, and inputs them along with the price movements of the preceding n-1 days into a time series forecasting model network. Finally, MOF forwards the results from the third step to a fully connected layer to predict stock price movements and output binary labels.


\subsection{System Design Overview}
Our forecasting system is designed as a multi-source fusion model that combines the advantages of large language models with traditional financial data analysis. The overall goal is to leverage language model to extract meaningful insights from textual data and integrate these insights with historical stock price data to improve the accuracy of forecasts.

\label{Sec.System Design Overview}



\subsubsection{Multi-Source Data Integration}
Our forecasting system is conceptualized as a multi-source fusion model that integrates a plethora of data types. At the core of this integration lies the amalgamation of the Large Language Model with traditional financial data analysis, thereby creating a comprehensive mechanism for stock price prediction.

\subsubsection{Linguistic Model Component}
The linguistic component of our model focuses on extracting key information from textual data to create a summary using a large language model. Subsequently, the MacBERT model processes and refines this policy summary information into MacBERT Feature I.

\subsubsection{Financial Data Processing}
Parallel to the linguistic processing, the mechanism also ingests Stock Reviews, which are similarly processed to yield Stock Review Summaries. These summaries are then converted into a set of features known as MacBERT Features II, capturing the sentiment and trends present in the stock market discourse.

\subsubsection{Fusion Mechanism}
The mechanis's innovative aspect lies in its fusion mechanism, where MacBERT Features I and MacBERT Features II are combined. This fusion facilitates a richer, multidimensional representation of the data, which is critical for accurate time series prediction of stock movements.

\subsubsection{Time Series Prediction Model}
The integrated features are then fed into a Neural Network, which serves as the backbone of our time series prediction model. This network encompasses a variety of deep learning architectures such as LSTM, BiLSTM, GRU, MogrifierLSTM, and ST-LSTM, each designed to capture different aspects of temporal dependencies within the stock price data.

\subsubsection{Prior Effect Incorporation}
To account for the influence of past stock movements on future trends, the model incorporates a Prior Effect module. This module enables the system to consider historical stock movements when making predictions, thus enhancing the model's ability to learn from sequential data.

\subsubsection{Output Layer Design}
The point of the model's processing occurs at the Fully Connected Output Layer, where the final predictions are made. This layer is tasked with translating the complex interactions of the integrated features and temporal dynamics into a binary output indicative of future stock movements.

This systematic design approach, with its multi-tiered architecture and sophisticated integration of linguistic and financial analytics, aims to push the boundaries of financial forecasting by providing a nuanced, data-driven perspective on stock price movements.

\subsection{Data Collection and Preprocessing}
\label{metapath}
Our approach begins with a well-designed data collection process, incorporating A-share market data such as the Shanghai and Shenzhen indices. Additionally, we gathered information on policies released by the People's Bank of China (PBOC) during the same period and utilized a large language model to generate summaries of these policies, providing insights into the macroeconomic context. We also collected legally obtained stock reviews from the same period, processed them using a large language model for future use. Data preprocessing is a critical step to ensure data standardization and readiness for subsequent analysis.

\subsubsection{Data parsing}
 Data parsing involves tasks such as date parsing, converting stock price movements into binary labels indicating price changes, and extracting summaries from policy information.
 
\subsubsection{ Data standardization}
To eliminate dimensional discrepancies across different datasets, we standardized stock price movement data using the following methods:

\textbf{Min-max normalization:} Scale the data to between 0 and 1, the formula is: 
$$x_{norm} =\frac{x-min(x)}{max(x)-min(x)} $$ \par

\textbf{Z-score normalization:} Normalized according to the mean ( \textbf{$\mu$}  ) and standard ( \textbf{$\sigma$} ) deviation of the data:
$$x_{std} =\frac{x-\mu }{\sigma } $$ \par

\subsubsection{Binary label conversion}
 Successive price changes are converted into binary labels, where 1 indicates a price increase and 0 indicates a price decrease. This provides a clear prediction target for the model.

 \subsubsection{Textual Feature Extraction}
Leveraging the MacBERT model, we process policy summaries to extract high-dimensional feature vectors. These vectors encapsulate deep semantic and sentiment information from the text, crucial for understanding the nuanced impact of policies on stock prices.

\subsubsection{Feature Vector Formation}
Through the encoding process, the MacBERT model transforms policy summary text into high-dimensional feature vectors. These vectors serve as distilled representations, capturing the essence of the policy's potential market impact.

\subsubsection{Advanced Masking Strategy}
A distinctive feature of MacBERT is its advanced masking strategy, enhancing the model's ability to handle Chinese language intricacies by substituting target words with semantically similar words during pre-training.

\subsubsection{Enhancing Model Interpretability}
Textual features extracted via MacBERT not only enhance our model's predictive power but also improve its interpretability. Stakeholders can gain insights into specific aspects of policy texts influencing predictions.

\subsection{Textual Feature Extraction }
During textual feature extraction using the MacBERT model, we delve into its internal mechanisms to accurately capture deep semantics and sentiment in policy texts.

\subsubsection{Application of the Pre-trained MacBERT Model}
The MacBERT model learns diverse linguistic features during pre-training, laying a solid foundation for subsequent fine-tuning tasks.  One of the pre-training tasks is the Masked Language Model (MLM), whose objective function can be represented as:
\begin{equation}
    \mathcal{L}_{\text{MLM}} = -\sum_{i=1}^{N} \log p(t_i)
\end{equation}
Where \( N \) is the number of masked tokens, \( t_i \) is the index of the \( i \)-th token, and \( p(t_i) \) is the model's predicted probability for that token.

\subsubsection{Semantic and Sentiment Information Capture}
The MacBERT model employs a bidirectional Transformer architecture, capturing relationships between words through the self-attention mechanism. The core of the self-attention mechanism is calculating the attention scores for each word with respect to all other words, as follows:
    \begin{equation}
        \text{Attention}(Q, K, V) = \text{softmax}\left(\frac{QK^T}{\sqrt{d_k}}\right)V
    \end{equation}
    Where \( Q \), \( K \), \( V \) represent the Query, Key, and Value matrices, respectively, and \( d_k \) is the dimension of the keys.

\subsubsection{Feature Vector Formation}
After processing by the encoder, each word is represented by an output vector \( h_i \), which encapsulates contextual information. The output vector \( h_{\text{start}} \) of the first word (usually a special start token) is used as the representation of the entire sentence for further processing.

\subsubsection{Advanced Masking Strategy}
The masking strategy employed by MacBERT during pre-training includes not only random masking but also the use of semantically similar words for masking, enhancing the model's understanding of the text. 

\subsubsection{Positional Encoding}
To enable the model to understand the order of words in a sequence, positional encoding is added to the input embeddings in MacBERT, with the formulas:
    \begin{equation}
        \text{PE}_{(pos, 2i)} = \sin(pos / 10000^{2i/d_{\text{model}}})
    \end{equation}
    \begin{equation}
        \text{PE}_{(pos, 2i+1)} = \cos(pos / 10000^{2i/d_{\text{model}}})
    \end{equation}
    Where \( pos \) is the position of the word, \( i \) is the dimension index, and \( d_{\text{model}} \) is the dimension of the model.

\subsubsection{Multi-Head Attention Mechanism}

MacBERT uses a multi-head attention mechanism to capture representations of information in multiple subspaces in parallel, with the formula:
    \begin{equation}
        \text{MultiHead}(Q, K, V) = \text{Concat}(\text{head}_1, \dots, \text{head}_h)W^O
    \end{equation}
    Where \( h \) is the number of heads, \( \text{head}_i \) is the attention output of the \( i \)-th head, and \( W^O \) is the output weight matrix.

\subsubsection{Layer Normalization and Residual Connections}
 To improve the stability and efficiency of model training, MacBERT uses layer normalization and residual connections in each sub-layer:
    \begin{equation}
        \text{LayerNorm}(x + \text{Sublayer}(L(x)))
    \end{equation}
    Where \( L(x) \) is the sub-layer (such as multi-head attention and feed-forward network), and \( \text{LayerNorm} \) is the layer normalization operation.

\subsubsection{Feed-Forward Network}
 Following the self-attention layer, MacBERT uses a feed-forward network to further process features, generally in the form of:
    \begin{equation}
        \text{FFN}(x) = \max(0, xW_1 + b_1)W_2 + b_2
    \end{equation}
    Where \( W_1 \), \( W_2 \) are weight matrices, and \( b_1 \), \( b_2 \) are bias terms.

\subsection{Model Architecture}

Our financial trend forecasting mechanism's model architecture is a symphony of layers designed to integrate textual features with stock price time series data effectively. Here, we outline each layer's role and its mathematical underpinning.

\subsubsection{Input Layer}
The model receives as input the feature vectors from the MacBERT model and the historical stock price data, represented as a sequence \( X = \{x_1, x_2, ..., x_t\} \), where \( x_t \) is the stock price at time \( t \).

\subsubsection{Embedding Layer}
Textual features are embedded into a higher-dimensional space for richer representation:
\[ E = W_e \cdot F + b_e \]
Here, \( W_e \) is the weight matrix, \( F \) is the original feature vector, and \( b_e \) is the bias vector.

\subsubsection{Convolutional Layer (For Text Data)}
We apply convolution to the embedded text features to extract local patterns:
\[ C = \text{Conv}(E) = f(W_c \star E + b_c) \]
The function \( f \) denotes an activation like ReLU, \( W_c \) and \( b_c \) are the convolutional layer's weights and biases.

\subsubsection{Recurrent Layer (LSTM/GRU)}
The recurrent layer, such as LSTM, captures temporal dependencies with update rules:
 \begin{align*}
i_t &= \sigma(W_i \cdot [h_{t-1}, x_t] + b_i) \\
f_t &= \sigma(W_f \cdot [h_{t-1}, x_t] + b_f) \\
\tilde{C}_t &= \tanh(W_C \cdot [h_{t-1}, x_t] + b_C) \\
C_t &= f_t \circ C_{t-1} + i_t \circ \tilde{C}_t \\
o_t &= \sigma(W_o \cdot [h_{t-1}, x_t] + b_o) \\
h_t &= o_t \circ \tanh(C_t)
\end{align*} 
The \( i_t \), \( f_t \), and \( o_t \) represent the input, forget, and output gates, respectively, and \( \circ \) denotes the Hadamard product.

\subsubsection{Attention Mechanism}
The attention mechanism assigns weights to different parts of the input sequence:
\[ \alpha_{tj} = \frac{\exp(\text{score}(h_{t-1}, F_j))}{\sum_{k}\exp(\text{score}(h_{t-1}, F_k))} \]
This allows the model to focus on relevant features for the prediction task.

\subsubsection{Fusion Layer}
The fusion layer combines textual and stock price data:
\[ Z = \gamma \cdot O_t + (1 - \gamma) \cdot C \]
The parameter \( \gamma \) balances the contributions from the output of the recurrent layer \( O_t \) and the convolutional layer \( C \).

\subsubsection{Fully Connected Layer}
The fused data is then passed through fully connected layers to perform classification:
\[ Y = W_f \cdot Z + b_f \]
Here, \( W_f \) and \( b_f \) are the weights and biases of the fully connected layer.

\subsubsection{Output Layer}
A sigmoid function in the output layer provides the probability of an event, such as a stock price increase:
\[ P = \sigma(Y) \]

\subsubsection{Loss Function}
The binary cross-entropy loss is used to train the model to predict accurately:
\[ \mathcal{L} = -\sum_{t} y_t \log(P_t) + (1 - y_t) \log(1 - P_t) \]
The \( y_t \) represents the actual target values and \( P_t \) the predicted probabilities.

\subsubsection{Backpropagation}
Finally, backpropagation calculates the gradients, and an optimizer like Adam updates the model parameters to minimize the loss.



\subsection{Experimental Design and Model Evaluation}
The following subsections provide a detailed account of our methodological rigor and empirical validation, showcasing the systematic steps undertaken to evaluate and substantiate the efficacy of our forecasting system.

\subsubsection{Data Partitioning and Sequence Formulation}
Our methodology begins with a strategic division of the dataset into training and testing subsets, adhering to an 80-20 ratio. This partitioning is pivotal for ensuring the robustness and generalizability of our model’s predictive capabilities. The formula for calculating the size of the training set based on the total data size \( T \) is:
\[ \text{Training Set Size} = 0.8 \times T \]

\subsubsection{Model Training and Optimization}
The training regimen utilizes the Adam optimizer, an adaptive learning rate optimization algorithm known for its efficiency in handling sparse gradients and noisy updates. We employ the binary cross-entropy loss function, which is particularly suited for binary classification tasks such as ours. The binary cross-entropy loss \( \mathcal{L}_{\text{BCE}} \) is given by:
\[ \mathcal{L}_{\text{BCE}} = -\left( \sum_{c=1}^{C} y_{o_c} \log(p_{c}) + (1 - y_{o_c}) \log(1 - p_{c}) \right) \]
Where \( C \) is the number of classes, \( y_{o_c} \) is the binary indicator (0 or 1) if class \( c \) is the correct classification for the observation, and \( p_{c} \) is the predicted probability that the observation is of class \( c \).

\subsubsection{Evaluation Metrics and Performance Benchmarking}
The efficacy of our model is rigorously evaluated using accuracy as the primary metric. Accuracy provides a quantitative measure of the model’s ability to correctly predict stock price movements, offering a straightforward assessment of its predictive prowess. The formula for accuracy \( A \) is:
\[ A = \frac{\text{Number of Correct Predictions}}{\text{Total Number of Predictions}} \]
To further validate our model’s superiority, we conduct a comparative analysis against a spectrum of established models, including but not limited to Neural Network, BiLSTM, LSTM, XGBoost, GRU, STLSTM, and MogrifierLSTM. This comparative analysis serves as a benchmark, highlighting our model’s enhanced prediction accuracy and interpretability.

\subsubsection{Comparative Analysis and Model Interpretability}
The comparative analysis not only benchmarks our model’s performance but also provides insights into its interpretability. By juxtaposing our model’s predictions with those of other models, we can deduce the extent to which the integration of linguistic features and temporal stock data contributes to the model’s predictive power. We also consider other metrics such as the F1 score, which combines the precision \( P \) and recall \( R \) into a single measure:
\[ F1 = 2 \times \frac{P \times R}{P + R} \]
This metric is particularly useful when the class distribution is imbalanced.

This integrated approach to experimental design and model evaluation ensures a comprehensive assessment of our forecasting system, positioning it as a robust and reliable tool for stock market analysis.
\section{Evaluation}
\label{sec:eval}
In this section, we provide a detailed evaluation of MOFs. We will first introduce our hardware setup. Next, we describe the dataset that we collected and processed. Following this, we focus on comparing the performance of various time series prediction models integrated into the mechanism, encompassing both traditional models and newer variants. Additionally, we present detailed case studies to facilitate thorough comparisons, further exploring and elucidating the role and potential of our system.


\subsection{Experimental environment}
The following describes the hardware devices and data set information we used in the evaluation process. Hardware information includes computer configuration, processor, memory and other key information. Data set information includes the data set we use, including data sources, data types, data preprocessing methods, etc.

\subsubsection{Experimental Configuration}
Our experimental computer was equipped with a ninth-generation Intel Core i7 processor and an NVIDIA GeForce GTX 1660 Ti stand-alone graphics card with 8GB of RAM.
Our experimental software setup includes PyTorch 2.0.0 and CUDA 11.7, running on a Windows 10 operating system. For regular and fully connected layers, we employed the Adam optimizer with a learning rate of 1e-3 and a batch size of 32.

\subsubsection{Dataset}We mainly collected A-share market data, including Shanghai Composite Index, Shenzhen Composite Index and China Securities index. Regarding policy information, we extracted policy summaries from the People's Bank of China (PBOC) using a large language model.

\begin{figure}[h]
	\includegraphics[width=80mm]{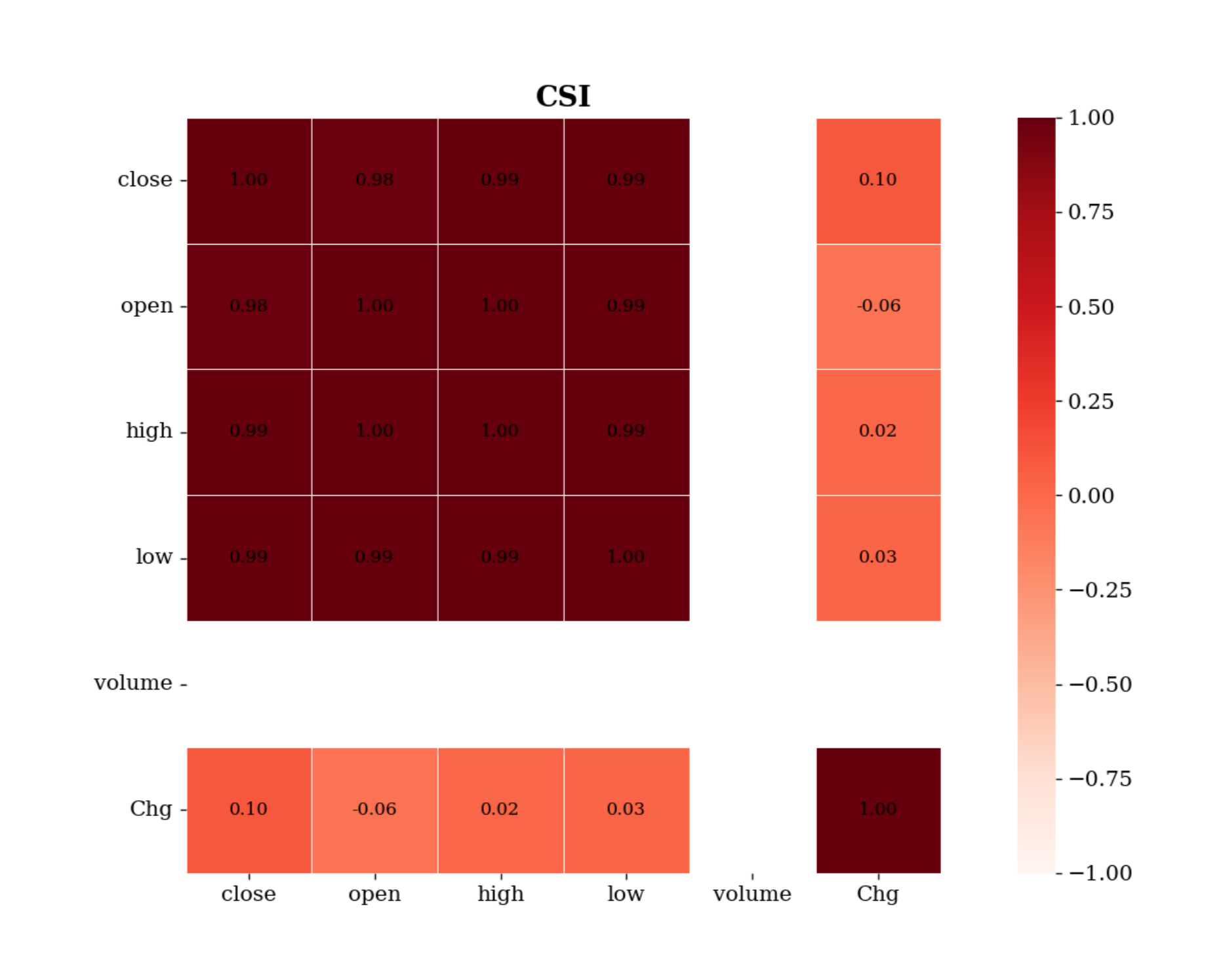}
	\centering
	\caption{CSI Heat Map}
	\label{fig:CSI}
\end{figure}

\begin{figure*}
    \centering
    \begin{subfigure}{0.49\textwidth} 
        \includegraphics[width=\linewidth]{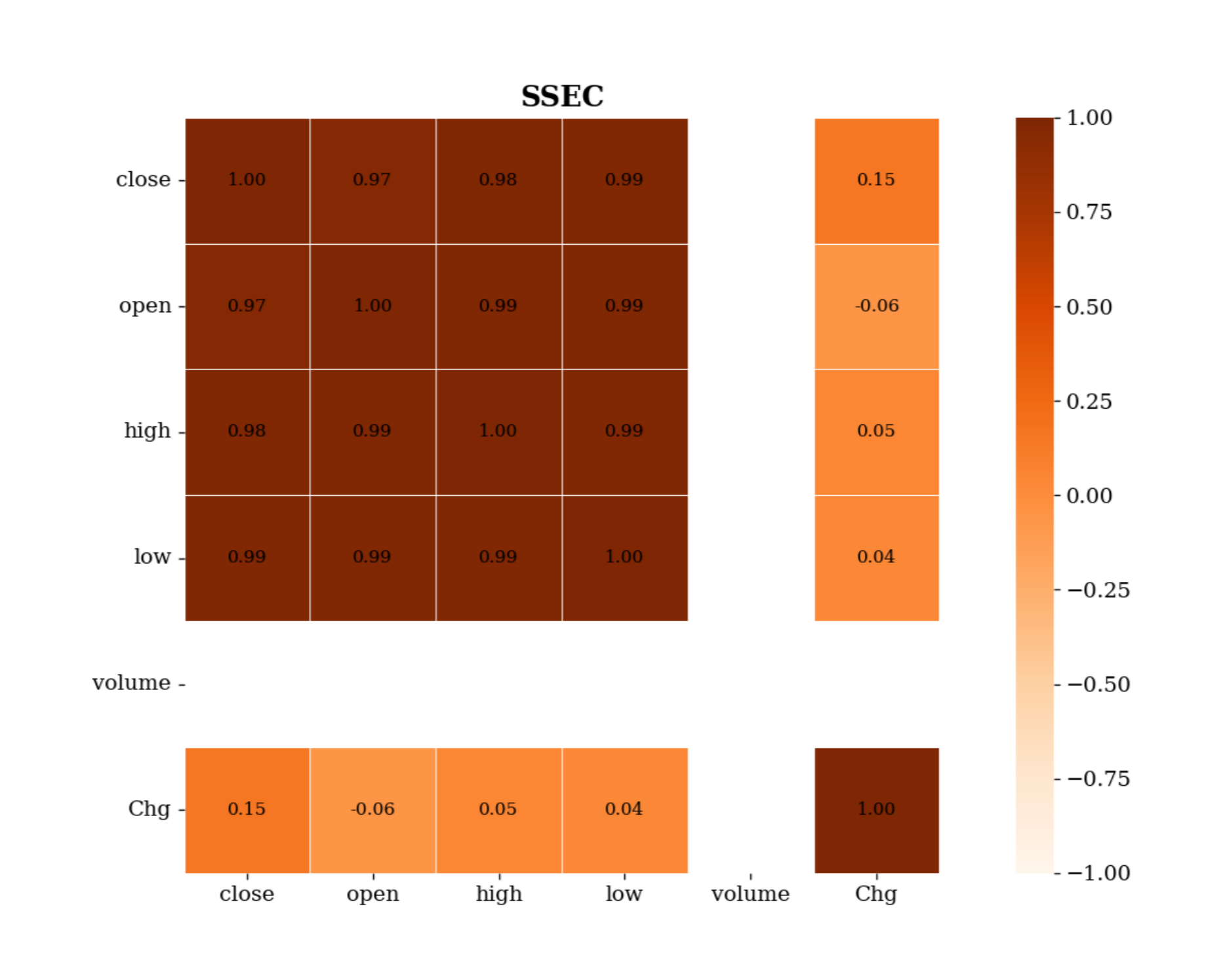}
        \caption{SSEC}
        \label{fig:SSEC}
    \end{subfigure}
    \hfill 
    \begin{subfigure}{0.49\textwidth} 
        \includegraphics[width=\linewidth]{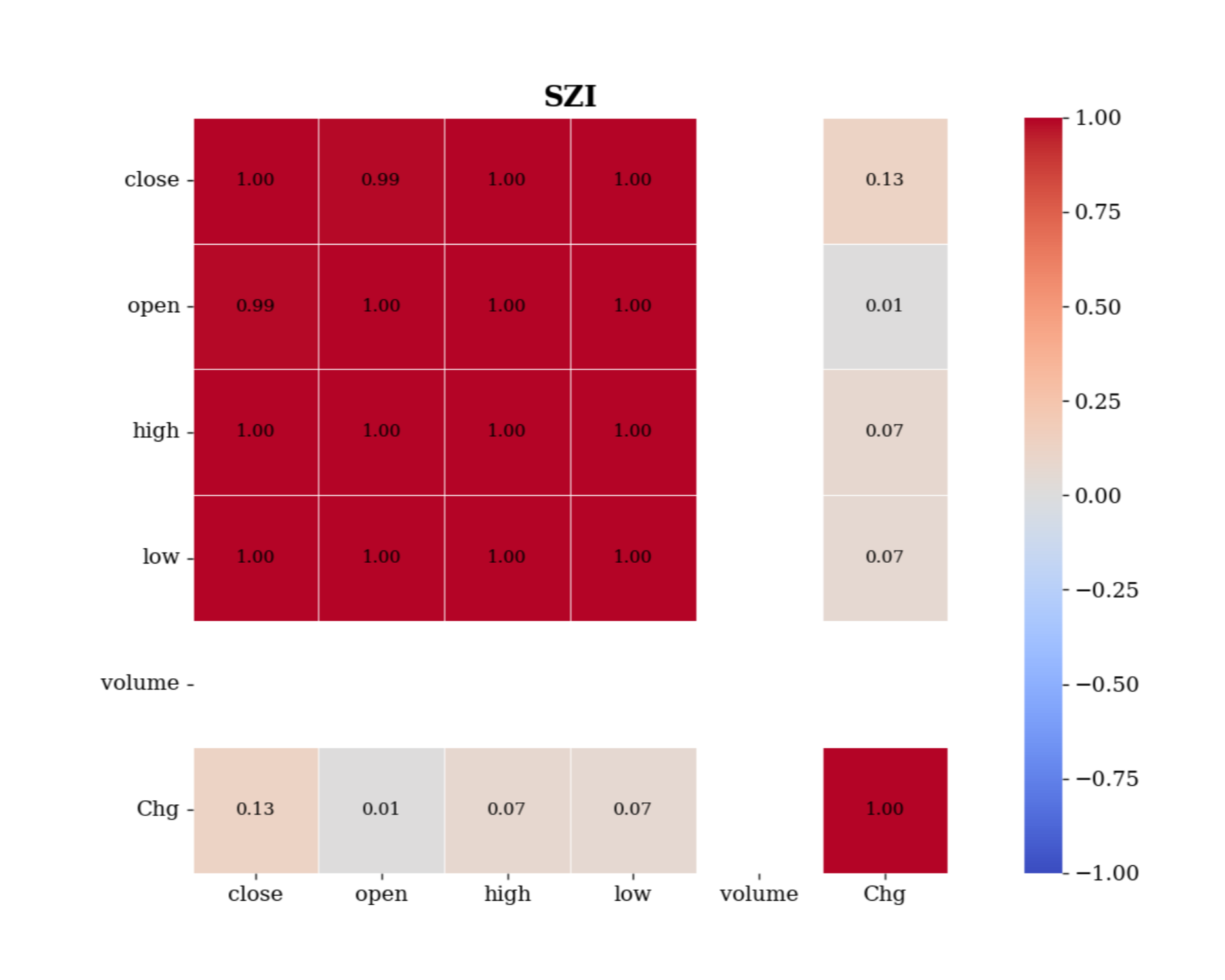}
        \caption{SZI}
        \label{fig:SZI}
    \end{subfigure}
    \caption{Stock Data Heat Map}
    \label{fig:combined}
\end{figure*}

\textbf{Stock Data:}We collected A-share market data from reliable financial information sources, including indices such as Shanghai Securities Composite Index (SSEC) , the Shenzhen Composite Index (SZI) and China Securities Index (CSI). These three indexes are important tools for investors to understand and analyze the Chinese stock market, and they each have different calculation methods and component compositions, which can show the dynamics and trends of the Chinese stock market from different angles.To facilitate visualization of the data sets, we drew their heat maps, as shown in Fig.~\ref{fig:CSI}, Fig.~\ref{fig:SSEC} and Fig.~\ref{fig:SZI}.

The data regarding stock price movements is derived from these three indexes, covering the period from December 2022 to December 2023. Through our own processing, we organize the data into a tabular format that includes date, change (Chg), opening price, closing price, trading volume, and other indicators. A data preprocessing step is applied to the stock price change information, which involves parsing the dates, converting the stock price change (Chg) into floating-point numbers, and categorizing them into binary labels (1 for increase, 0 for decrease).

\textbf{Policy Information:}
We obtained policy summaries from the People's Bank of China (PBOC) through a legitimate web crawler or existing database. Text features are extracted from these summaries, where non-string values are replaced with NaN. The textual features are represented as multiple high-dimensional matrices. To ensure consistency in subsequent prediction processes, we standardized these text features by reducing the dimensionality of the high-dimensional matrices to one dimension. We then filled or truncated them to a fixed length to ensure uniformity across the dataset.
\subsubsection{Evaluation Index}
We used the four metrics most commonly used in machine learning-related research, namely accuracy, F1, precision value, and recall rate.

\textbf{Accuracy:} Accuracy is the proportion of the number of correctly classified samples to the total number of samples, and it is an intuitive indicator. Accuracy is calculated as follows:
$$\text{Accuracy} = \frac{ \text{ Correct Sample } }{\text{Sample Count}}\times 100 \% $$\par
\textbf{Precision:} Accuracy is the proportion of positive categories that the model predicts (such as a rising stock) are actually positive, and a high accuracy means that the model rarely misclassifies negative categories into positive ones.Precision is calculated as follows:
$$\text{Precision} =\frac{ \text{  TP } }{\text{ TP}+\text{ FP }} \times 100 \% $$\par
\textbf{Recall:} The recall rate, also known as the true rate or sensitivity, is the proportion of samples that are actually in the positive category that the model correctly predicts to be in the positive category. The high recall rate means that the model is able to capture most of the positive class samples.Precision is calculated as follows:
$$\text{Recall} =\frac{ \text{  TP } }{\text{ TP}+\text{ FN }} \times 100 \% $$\par\par
\textbf{F1 Score:} The F1 score is a harmonic average of accuracy and recall, and it attempts to consider both metrics at the same time. F1 scores are balanced between accuracy and recall and are an indicator that takes both into account. F1 is calculated as follows:
$$\text{F1 Score} = 2 \times \frac{\text { Recall } \times \text { Precision }}{\text { Recall }+\text { Precision }}$$\par




\subsection{Time Series Prediction Model}
We embed eight models that can be used for time series prediction into the mechanism, fully compare the performance of each model, and analyze the advantages of each model:

\subsubsection{Neural Network}
In time series prediction, neural networks are renowned for their robust nonlinear modeling capabilities, automatic feature extraction, ability to process multi-variable inputs, capture sequence dependencies, and exhibit strong generalization ability and scalability. Despite their higher computational demands and data volume requirements, along with risks of overfitting and interpretability challenges, neural networks, when properly trained, are effective for time series prediction tasks, providing accurate and reliable results.
\subsubsection{Long Short Term Memory Network }
LSTM (Long Short-Term Memory Network) stands out as a specialized type of recurrent neural network that addresses the issues of gradient vanishing or exploding in traditional RNNs through gating mechanisms. It is particularly well-suited for time series prediction due to its capability to capture long-term dependencies, autonomously learn and extract critical features, handle nonlinear relationships within sequence data, and demonstrate strong generalization ability. Moreover, LSTM's capacity to handle sequences of varying lengths enhances its flexibility and efficacy in time series forecasting.
\subsubsection{Bidirectional Long Short Term Memory Network }
BiLSTM (Bidirectional Long Short-Term Memory Network) enhances its capability to capture long-term dependencies and intricate patterns in time series prediction by considering both preceding and succeeding information in the sequence. This approach significantly improves prediction accuracy and robustness, albeit at the cost of increased computational resources and data requirements.
\subsubsection{eXtreme Gradient Boosting }
XGBoost (eXtreme Gradient Boosting) is renowned in time series prediction for its efficient gradient boosting algorithm, exceptional ability to handle nonlinearity and large feature sets, and effective regularization techniques to prevent overfitting. It supports custom functions and parallel computation, thereby enhancing model performance and accelerating training speed.
\subsubsection{Mogrifier Long Short Term Memory Network }
MogrifierLSTM (Mogrifier Long Short-Term Memory Network) is a variant derived from LSTM that augments the expressive capacity of the LSTM network by introducing additional parameters. This adaptation enables MogrifierLSTM to effectively capture more complex patterns inherent in time series data with highly intricate and nonlinear characteristics, thereby improving prediction accuracy.
\subsubsection{Gated Recurrent Unit}
Gated Recurrent Unit (GRU) is a simplified version of LSTM that controls the flow of information through a gating mechanism. It captures long-term dependencies in time series data while having fewer parameters and a simpler structure, which generally makes GRUs faster to train and reduces the risk of overfitting. GRU is particularly suitable for processing time series data with fewer long-term dependencies and can be comparable to LSTM in some cases, but is more efficient in terms of the number of parameters and computational complexity.

\subsubsection{Spatio-Temporal Long Short-Term Memory}
ST-LSTM (Spatio-Temporal Long Short-Term Memory) is a variant of LSTM designed for spatio-temporal data. By integrating spatial and temporal features, it improves the ability to capture complex dynamics, making it especially suitable for multi-dimensional time series prediction tasks, although it requires more computational resourc

\subsubsection{Swin Transformer Long Short-Term Memory}
SwinLSTM (Swin Transformer Long Short-Term Memory) is a neural network unit that combines the self-attention mechanism of the Swin Transformer with the recurrent structure of LSTM. It captures global spatial dependencies through the self-attention mechanism and maintains long-term temporal dependencies through the gated unit of LSTM. This combination enhances the expressiveness and prediction accuracy of the model in time series prediction, making it especially suitable for processing sequence data with complex temporal and spatial characteristics.

\subsection{Experimental Results}
We first compared the accuracy of six models (LSTM, BiLSTM, MogrifierLSTM, GRU, ST-LSTM, SwinLSTM) embedded in MOF for predicting the rise and fall of stocks in the China Securities Index dataset, as shown in Table 1.The results indicate that, in our system, due to the inclusion of policy information and other influential characteristics, the accuracy of predicting stock movements is generally favorable.

\begin{table*}[t]
	\centering
	\caption{Experimental Results on China Securities index}
	\label{table:cwe_result1}
 \renewcommand\arraystretch{1.5}
 \tabcolsep=0.68cm
\begin{tabular}{cccccccccccc}
\hline
\multirow{2}{*}{\textbf{Models}} & \multicolumn{2}{c}{100ep}                         & \multicolumn{2}{c}{200ep}                         & \multicolumn{2}{c}{400ep}                         \\ \cline{2-7} 
                                 & \multicolumn{1}{c}{Acc} & \multicolumn{1}{c}{Pre} & \multicolumn{1}{c}{Acc} & \multicolumn{1}{c}{Pre} & \multicolumn{1}{c}{Acc} & \multicolumn{1}{c}{Pre} \\ \hline
LSTM                             & 0.7333                  & 0.7501                  & 0.6334                  & 0.6667                  & 0.6667                  & 0.6875                  \\
BiLSTM                           & 0.7241                  & 0.7692                  & 0.6896                  & 0.7143                  & 0.6897                  & 0.7142                  \\
MogrifierLSTM                    & 0.6897                  & 0.6667                  & 0.7241                  & 0.7692                  & 0.5862                  & 0.6001                  \\
GRU                              & 0.5517                  & 0.5385                  & 0.5517                  & 0.5556                  & 0.6897                  & 0.7143                  \\
ST-LSTM                          & 0.6552                  & 0.6471                  & 0.7931                  & 0.9091                  & 0.7586                  & 0.8333                  \\
SwinLSTM                         & 0.7667                  & 0.8001                  & 0.6667                  & 0.7501                  & 0.6666                  & 0.7501                  \\ \hline
\end{tabular}
\end{table*}

\begin{itemize}
\item \textbf{Multi-source data fusion enhances the prediction ability of the model.} 
The MOF constructs a multi-dimensional data set by integrating stock price data, policy text summaries, and stock commentary summaries. The tabular data shows that the accuracy rate of ST-LSTM on the 400-epoch window reaches 0.7931, with a precision rate of 0.9091. This indicates that multi-source data fusion significantly improves the model's prediction accuracy of stock price changes. This integration not only increases the richness of the data but also allows the models to capture more complex market dynamics and policy implications.

\item \textbf{The model structure design improves the depth and breadth of time series prediction.}  
The MOF employs deep learning architectures such as LSTM and its variants, which are particularly well-suited for working with time series data and capturing long-term dependencies. The high accuracy of ST-LSTM and SwinLSTM in the table shows that the time series prediction layer of the model can effectively learn from stock price data and predict future trends. Additionally, the fully connected output layer of the model is able to translate complex temporal dynamics and text features into predictions, further enhancing the depth and breadth of the model's predictions.

\item \textbf{Optimization strategies ensure the efficiency and robustness of model training.}  
The Adam optimizer and the binary cross-entropy loss function mentioned in this paper provide an effective training strategy for the model. The tabular data shows that the accuracy and precision rates of all models are generally higher across different time windows, reflecting the effectiveness of the optimization strategy. Additionally, hyperparameter tuning further ensures the robustness of the model during training and helps avoid overfitting.
\item \textbf{The interpretability of the model provides transparency to investment decisions.} 
The MOF model not only performs well in prediction accuracy but also enhances interpretability by using the text features extracted by the MacBERT model. This allows investors and analysts to better understand how the model makes predictions based on policy texts and market sentiment. The differences in performance among the various models in the table can help researchers identify which features are most critical to the forecast, thereby improving transparency and trust in investment decisions.
\end{itemize}

We compared five time series prediction models embedded in the MOF mechanism: LSTM, BiLSTM, MogrifierLSTM, ST-LSTM, and SwinLSTM. We conducted stock rise and fall forecasting training on two datasets, the Shenzhen Index and the Shanghai Index. Training was conducted for 100, 200, and 400 rounds respectively, and the precision for each round corresponding to each dataset was recorded, as shown in Table 2. The results indicate that, due to the inclusion of influential factors such as policy information, the precision for these two datasets is quite high in our mechanism.

Focusing on the data and models in Table ~\ref{table:cwe_result}, the results highlight the predictive power of the MOF system under different time windows and market conditions. The ST-LSTM model exhibits the highest precision of 0.9091 in a 400-epoch time window, demonstrating its strong ability to capture long-term dependencies. The MogrifierLSTM model shows stability in the 200-epoch and 400-epoch windows with precisions of 0.8462 and 0.8182, respectively, supporting the robustness of the system. The LSTM model achieved a perfect precision of 0.9999 in the 100-epoch window and maintained high precision in the longer windows, showing its ability to quickly adapt to market changes. The BiLSTM model's high precision of 0.9999 in the 400-epoch window highlights its advantages in long-term forecasting. Additionally, the high precision of the ST-LSTM model, 0.9090 and 0.9091, on different market indices (SZI and SSEC) proves that the MOF mechanism can adapt to the characteristics of different market indices and provide customized forecasting services. Together, these findings highlight the advantages of the MOF mechanism in terms of precision, robustness, adaptability, and applicability, providing investors with a reliable stock market prediction tool.

\begin{table*}[]
	\centering
	\caption{Precision Comparison of Shenzhen Composite Index and Shanghai Composite Index Predictions}
	\label{table:cwe_result}
  \renewcommand\arraystretch{1.5}
 \tabcolsep=0.68cm
\begin{tabular}{cccccccccccc}
\hline
\multirow{2}{*}{\textbf{Models}} & \multicolumn{2}{c}{100ep}                         & \multicolumn{2}{c}{200ep}                         & \multicolumn{2}{c}{400ep}                         \\ \cline{2-7} 
                                 & \multicolumn{1}{c}{SZI} & \multicolumn{1}{c}{SSEC} & \multicolumn{1}{c}{SZI} & \multicolumn{1}{c}{SSEC} & \multicolumn{1}{c}{SZI} & \multicolumn{1}{c}{SSEC} \\ \hline
LSTM                             & 0.9999                  & 0.9167                  & 0.8333                  & 0.9167                  & 0.8889                  & 0.8571                  \\
BiLSTM                           & ------                 & ------                  & 0.9999                  & 0.9999                  & 0.9999                  & 0.9167                  \\
MogrifierLSTM                    & 0.9000                  & 0.9999                  & 0.8462                  & 0.8667                  & 0.8182                  & 0.9999                  \\

ST-LSTM                          & 0.9999                  & ------                  & 0.9090                  & 0.9999                  & 0.9091                  & 0.9999                  \\
SwinLSTM                         & ------                  & ------                  & 0.9999                  & 0.8000                  & 0.8000                  & 0.7500                  \\ \hline
\end{tabular}
\end{table*}

The following conclusions can be drawn:

\begin{itemize}
\item \textbf{Effectiveness of Multi-Source Data Fusion is Optimistic.} 
The model performance presented in Table ~\ref{table:cwe_result} indicates that the fusion of multi-source data, such as stock price data, policy text summaries, and stock review summaries, can significantly improve prediction accuracy. This supports the effectiveness of the multi-source data integration approach in the design of the MOF mechanism and emphasizes the importance of considering various information sources in financial trend forecasting.

\item \textbf{MacBERT Model's Strong Capability for Text Feature Extraction.}  
The MOF mechanism utilizes text features extracted by the MacBERT model, which play a crucial role in enhancing prediction accuracy. The high accuracy results in Table~\ref{table:cwe_result} further confirm MacBERT's ability to capture deep semantic and sentiment information from policy texts and market sentiment.

\item \textbf{High Applicability of Time Series Prediction Models in MOF.}  
The prediction results for different time windows in Table~\ref{table:cwe_result} demonstrate the applicability and effectiveness of LSTM and its variants (such as ST-LSTM) in time series prediction tasks. This supports the decision of the MOF mechanism to adopt these deep learning architectures for its time series prediction layer.

\item \textbf{Positive Impact of Optimization Strategies on Model Performance.} 
The MOF mechanism employs the Adam optimizer and binary cross-entropy loss function, combined with hyperparameter tuning, to ensure efficient and robust model training. The high accuracy performance of different models across various time windows in Table 2 demonstrates the success of these optimization strategies in preventing overfitting and improving model generalization.
\item \textbf{Generalization Ability of the Models.} 
The data in~\ref{table:cwe_result} indicate that, despite fluctuations in performance across different time windows, most models maintain high precision consistently. This reflects the MOF mechanism's strong generalization ability under varying market conditions, providing investors with stable prediction results.

\end{itemize}

We also compared the training losses of six time series prediction models embedded in the MOF mechanism using the CSI dataset. These models include LSTM, BiLSTM, GRU, MogrifierLSTM, ST-LSTM, and SwinLSTM. We recorded the loss values for each training epoch, plotted them as scatter plots, and conducted a comparative analysis. Combining these findings with the innovations of our MOF mechanism, we derived the following insights:

\begin{itemize}
\item \textbf{The advantages of multi-source data fusion are evident.} 
The MOF mechanism constructs a multi-dimensional dataset by integrating stock price data, policy text summaries, and stock commentary summaries. The loss graph in Fig~\ref{fig:CSI}.  shows that the model's accuracy in predicting stock fluctuations is generally optimistic due to the inclusion of influential features such as policy information. This indicates that multi-source data fusion significantly enhances the model's accuracy in predicting stock price changes.

\item \textbf{The interpretability of our system improves the training effectiveness for predicting stock movements.}  
The MOF mechanism not only achieves high prediction accuracy but also enhances model interpretability by leveraging text features extracted by the MacBERT model. This transparency allows stakeholders to better understand how the model uses policy texts and market sentiment in making predictions, thereby improving the effectiveness of training for stock movement predictions.

\item \textbf{The MOF system demonstrates strong generalization capabilities.}  
As depicted in Fig~\ref{fig:CSI}, the loss for all models gradually decreases during the training process, indicating that the models effectively learn from the training data and enhance their predictive abilities. Moreover, the consistent reduction in loss demonstrates their robust generalization capabilities, enabling accurate predictions on unseen data.
\end{itemize}



\begin{figure*}[htbp] 
    \centering
    \begin{subfigure}{0.49\textwidth} 
        \includegraphics[width=\linewidth]{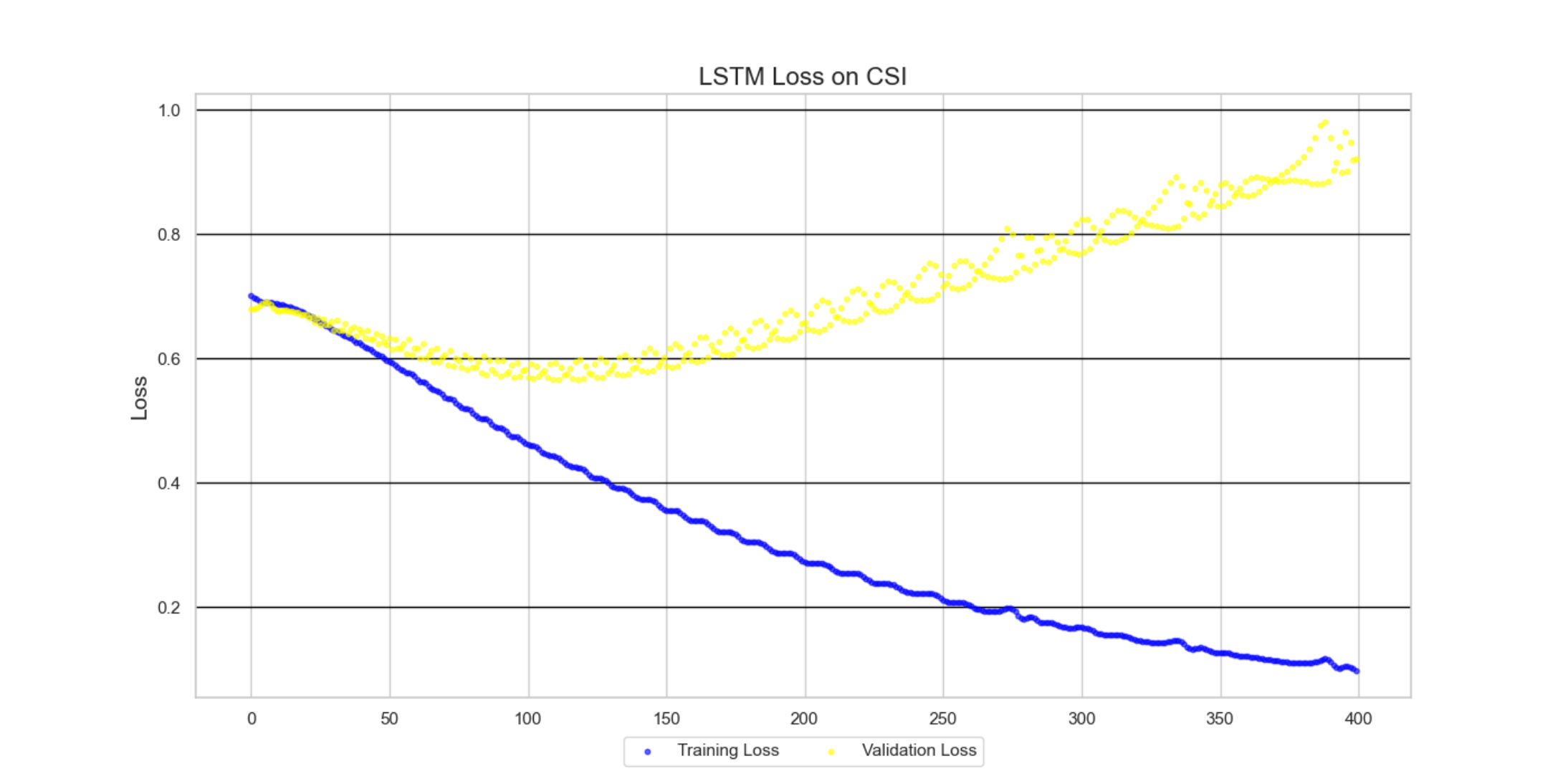}
        \caption{LSTM Loss}
        \label{fig:LSTM}
    \end{subfigure}
    \hfill 
    \begin{subfigure}{0.49\textwidth} 
        \includegraphics[width=\linewidth]{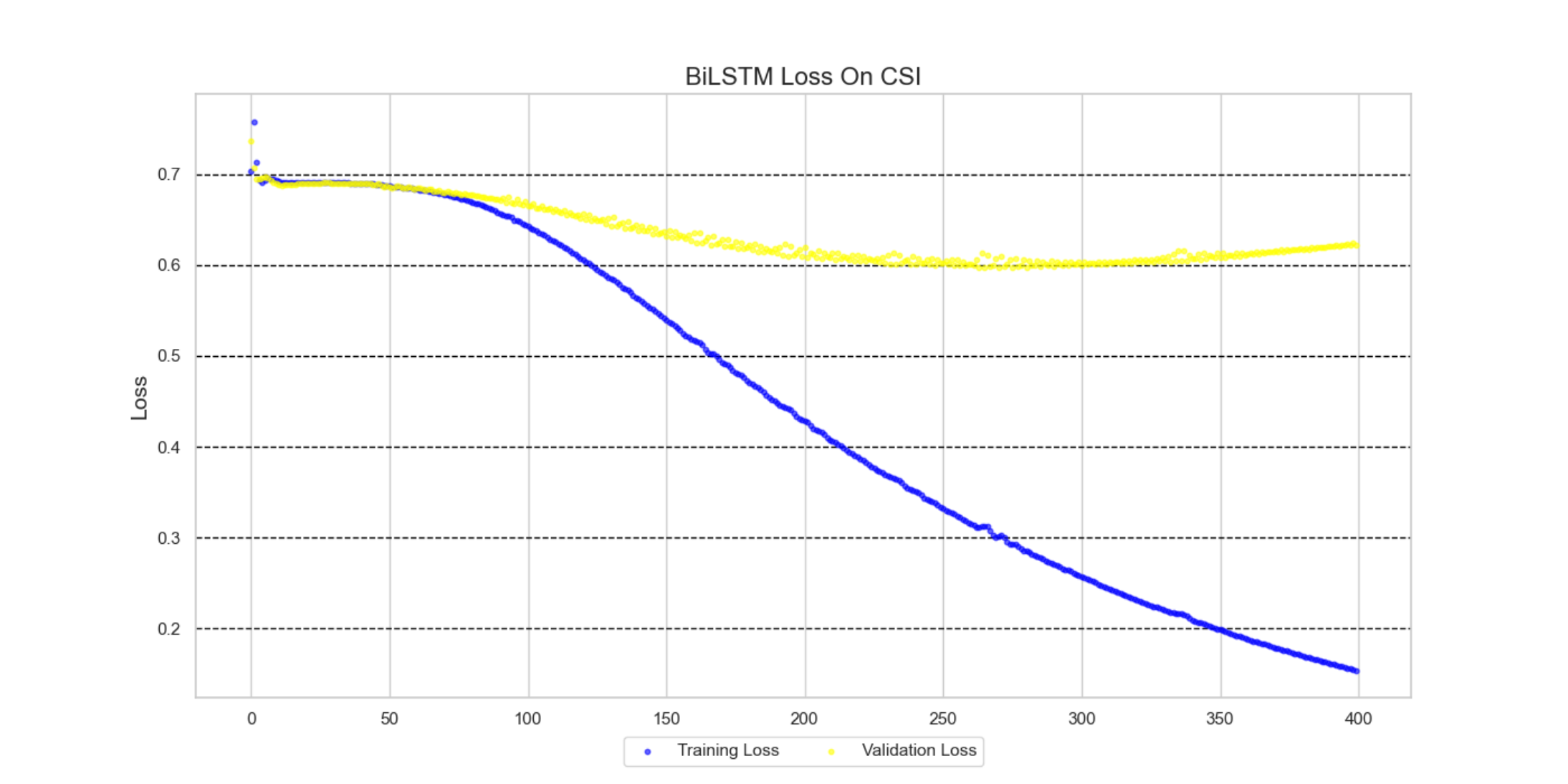}
        \caption{BiLSTM Loss}
        \label{fig:BiLSTM}
    \end{subfigure}

    \begin{subfigure}{0.49\textwidth} 
        \includegraphics[width=\linewidth]{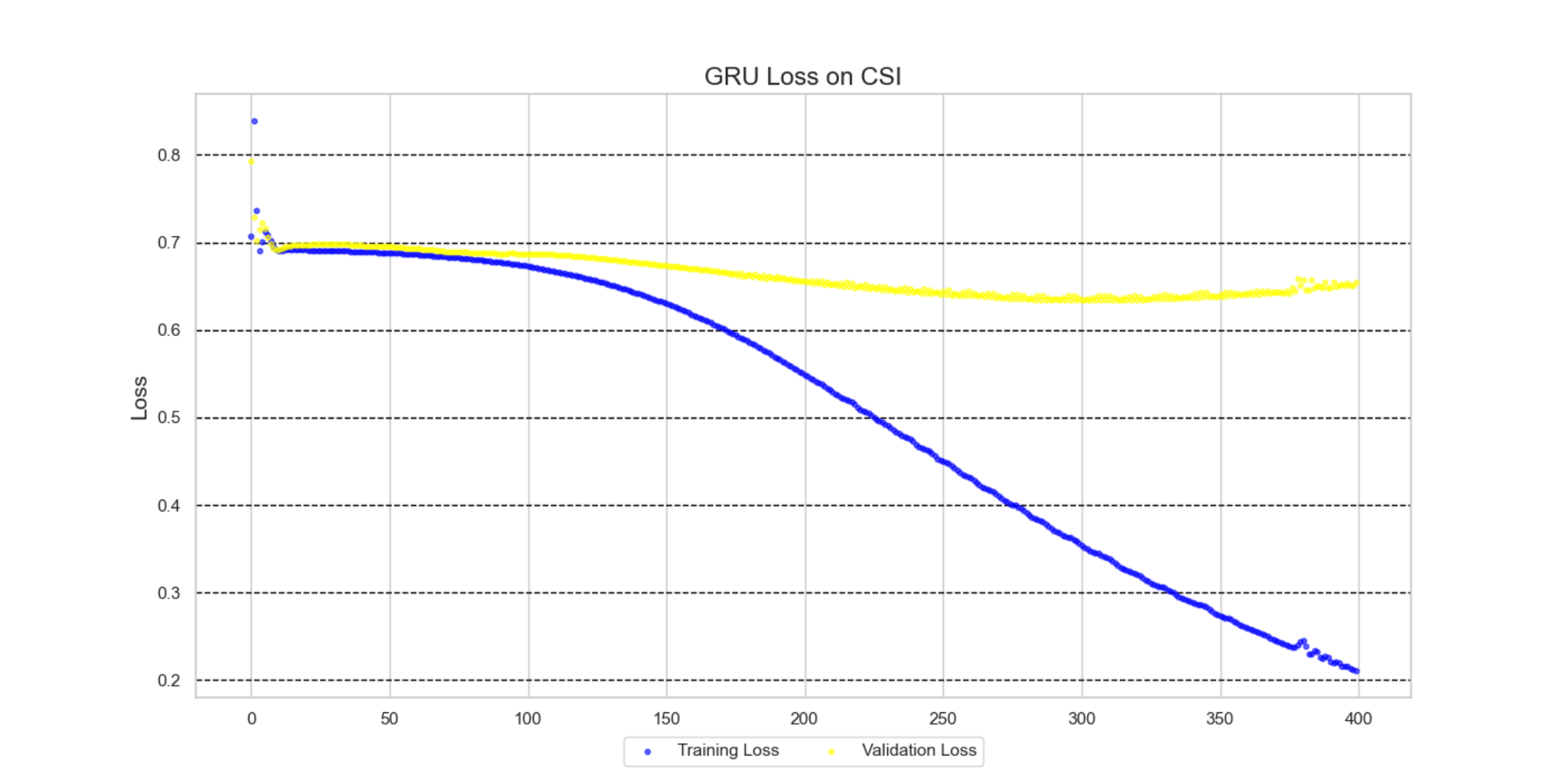}
        \label{fig:LSTM}
    \end{subfigure}
    \hfill 
    \begin{subfigure}{0.49\textwidth} 
        \includegraphics[width=\linewidth]{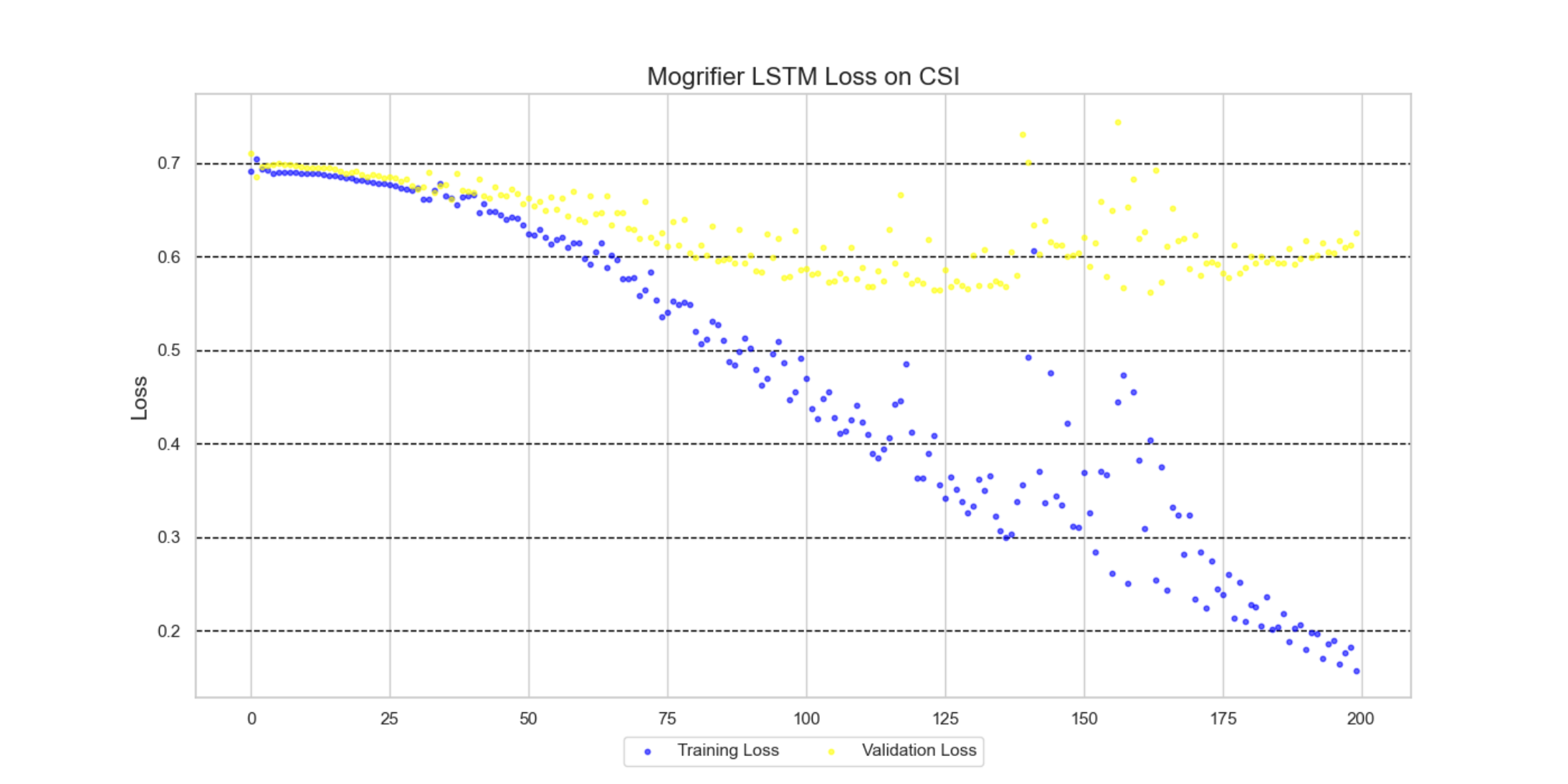}
        \caption{MogrifierLSTM Loss}
        \label{fig:BiLSTM}
    \end{subfigure}

    \begin{subfigure}{0.49\textwidth} 
        \includegraphics[width=\linewidth]{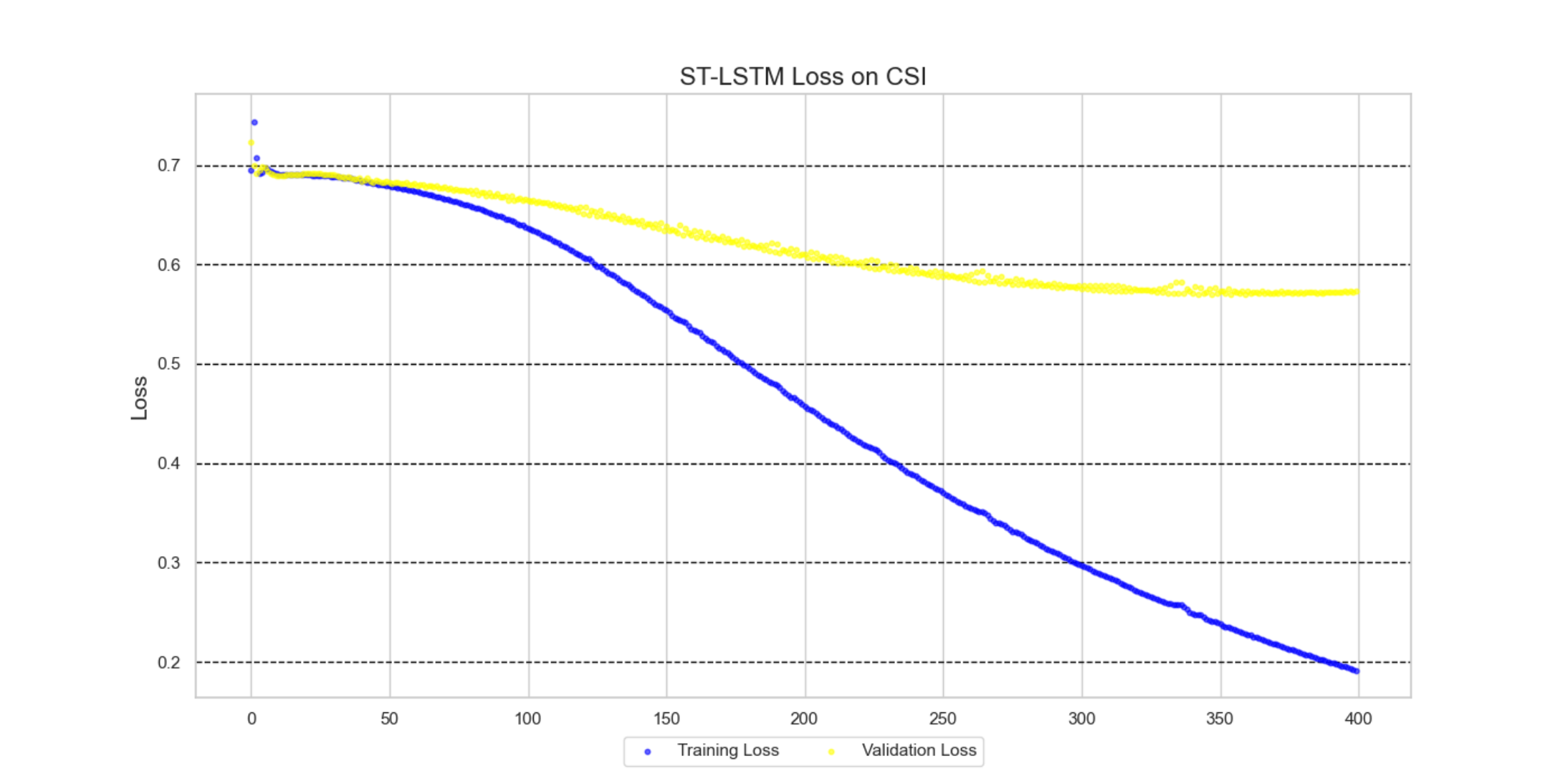}
        \caption{ST-LSTM Loss}
        \label{fig:LSTM}
    \end{subfigure}
    \hfill 
    \begin{subfigure}{0.49\textwidth} 
        \includegraphics[width=\linewidth]{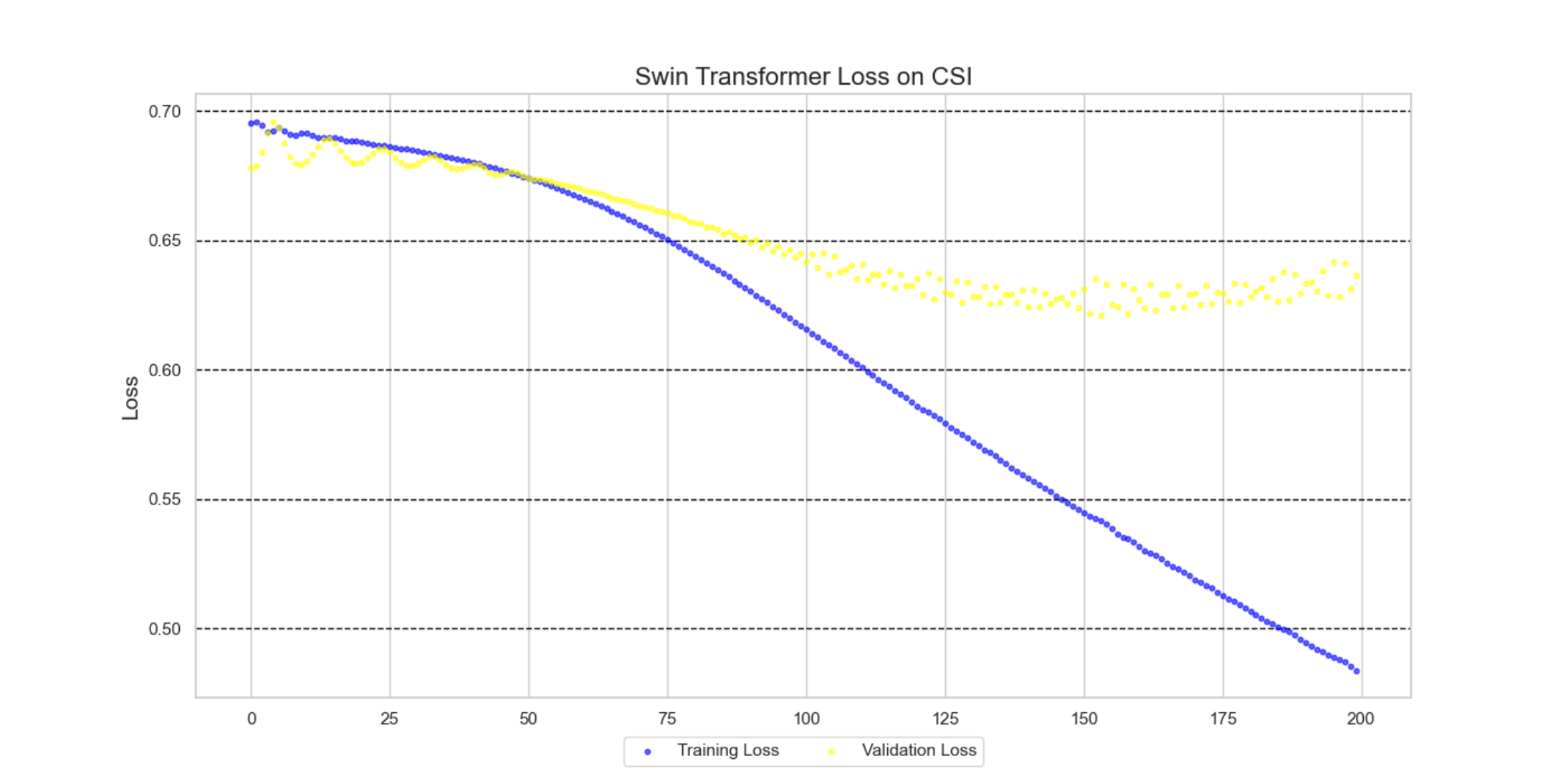}
        \caption{SwinLSTM Loss}
        \label{fig:BiLSTM}
    \end{subfigure}
    
    \caption{Comparison of Loss on CSI}
    \label{fig:combined}

\end{figure*}

\subsection{Ablation Study}
We conducted an ablation study on the two embedding models within this mechanism, examining their performance with and without the "Prior Effect". The "Prior Effect" represents the stock's rise and fall in the previous N-1 days, encoded as a one-dimensional matrix consisting of 1s and 0s, where 1 indicates a rise and 0 indicates a decline. The results demonstrate that the "Prior Effect" positively influences the prediction accuracy of stock rise and fall.

\textbf{Previous effects have a positive effect on the predictive power of the model.} The ablation study reveals that the "Prior Effect" enhances the prediction of stock fluctuations. Specifically, without considering the "Prior Effect", the Neural Network achieved an F1 score of 0.6177 and a recall rate of 0.5055. With the inclusion of the "Prior Effect", the F1 score improved to 0.7428, and the recall rate increased to 0.8125. Similarly, for the LSTM model, the F1 score increased from 0.6451 with a recall of 0.6251 to 0.7692 with a recall of 0.6667. This underscores that incorporating the historical rise and fall of stocks over the past N-1 days significantly enhances the model's accuracy in predicting future stock trends.

\textbf{The explanatory power of the model is enhanced by the features extracted by the MacBERT model.} The MOF mechanism utilizes text features extracted by the MacBERT model, enhancing both prediction accuracy and model interpretability. This enables stakeholders to gain deeper insights into how predictions are derived from policy texts and market sentiment. In ablation studies comparing model performance with or without the "Prior Effect," we can further elucidate the reasoning behind model predictions, thereby enhancing transparency in investment decisions.

\textbf{Optimization strategies ensure the efficiency and robustness of model training.} The Adam optimizer and the binary cross entropy loss function mentioned in this paper provide an effective training strategy for the model. Data from the ablation study showed that the model maintained high accuracy and accuracy rates over different time Windows with or without Prior Effect, reflecting the effectiveness of the optimization strategy. In addition, hyperparameter tuning further ensures robustness during model training and helps to avoid overfitting.


\begin{table}[htbp]
	\centering
	\caption{Ablation Study}
	\resizebox{\columnwidth}{!}{
	\label{table:ablation}
 \renewcommand\arraystretch{1.5}
	\begin{tabular}{clllllllll}
\hline
\multicolumn{2}{c}{\multirow{2}{*}{\textbf{Method}}} & \multicolumn{4}{c}{Without Prior Effect}                & \multicolumn{4}{c}{With Prior Effect}                   \\ \cline{3-10} 
\multicolumn{2}{c}{}                                 & \multicolumn{2}{c}{F1}     & \multicolumn{2}{c}{Recall} & \multicolumn{2}{c}{F1}     & \multicolumn{2}{c}{Recall} \\ \hline
\multicolumn{2}{c}{Neural Network}                   & \multicolumn{2}{l}{0.6177} & \multicolumn{2}{l}{0.5055} & \multicolumn{2}{l}{0.7428} & \multicolumn{2}{l}{0.8125} \\
\multicolumn{2}{c}{LSTM}                             & \multicolumn{2}{l}{0.6451} & \multicolumn{2}{l}{0.6251} & \multicolumn{2}{l}{0.7692} & \multicolumn{2}{l}{0.6667} \\ \hline
\end{tabular}
	}
\end{table}

\section{Related Works}
\label{sec:related}

The research on stock price prediction be divided into three categories, i.e., traditional methods, machine learning-based methods, and deep learning-based methods.

\textbf{Machine Learning-based Methods.} 
Dey\emph{et al.} used the XGBoost algorithm to design an effective model to predict stock trends using technical indicators as features~\cite{dey2016forecasting}. Di Persio\emph{et al.} compare three different RNNS architectures, i.e., basic RNN, LSTM, and gated recursive unit (GRU) to assess which RNN performs better in predicting Google's stock price movements~\cite{di2017recurrent}. Leung\emph{et al.} connect collaborating companies in the information technology sector in a graph structure and use an SSVM to predict positive or negative movement in their stock prices~\cite{leung2014machine}. Xiao \emph{et al.} propose the cumulative autoregressive moving average method combined with the least squares support vector machine synthetic model (ARI-MA-LS-SVM) for basic stock market forecasting. Secondly, the data processing of predictive indicators is firstly carried out using cumulative autoregressive moving averages. Then, a least squares support vector machine using a simple indicator system is used to predict stock price fluctuations~\cite{xiao2020stock}. Kohara \emph{et al.} used a priori knowledge to extract event information and economic indicators, combined with a neural network model for multivariate stock price prediction, and the experimental results showed the high prediction accuracy and effectiveness of the method~\cite{kohara1997stock}. De Fortuny \emph{et al.} evaluated the applicability of different performance metrics by designing stock price prediction models based on text mining techniques, and explored how to evaluate, validate, and improve these models in real-world applications with the aim of improving the accuracy of stock price prediction.~\cite{de2014evaluating}.  Soni\emph{et al.}explore different techniques used to predict stock prices, from traditional machine learning and deep learning methods to neural networks and graph-based approaches~\cite{soni2022machine}.Vijh\emph{et al.} used artificial neural networks and random forest techniques to predict the next day's closing prices of five companies belonging to different industries  ~\cite{vijh2020stock}.Zhang\emph{et al.} proposed a stock price trend prediction system that uses an unsupervised heuristic algorithm to predict stock price movements within a predetermined prediction time frame~\cite{zhang2018novel}. 

\textbf{Deep Learning-based Methods.} 
Yang\emph{et al.} introduced a method that integrates deep neural networks to model and forecast Chinese stock market indices, using recent indices as inputs~\cite{yang2017stock}.  Lu \emph{et al.} proposed a CNN-BiLSTM-AM approach to predict the next day's stock closing price. The technique consists of an Attention Mechanism (AM), a Bidirectional Long Short Term Memory (BiLSTM), and Convolutional Neural Network (CNN) for extracting features from the input data~\cite{lu2021cnn}. Islam \emph{et al.}conducted a comparative study on stock price prediction using three different methods: autoregressive integrated moving average, artificial neural network, and stochastic process-geometric Brownian motion ~\cite{islam2020comparison}. Yu \emph{et al.} designed a deep neural network prediction model to forecast stock prices using the Phase Space Reconstruction (PSR) method and LSTM~\cite{yu2020stock}. Selvin \emph{et al.}  predict the stock prices of businesses listed on the NSE using three different deep learning architectures and compare their performance~\cite{selvin2017stock}. Jin \emph{et al.} 
propose a stock market prediction method that incorporates an attention mechanism into Long Short-Term Memory (LSTM) networks, applies Empirical Mode Decomposition (EMD) to break down time series, and fully accounts for investor sentiment~\cite{jin2020stock}. Based on the Long Short-Term Memory (LSTM) deep learning algorithm, Kim-Sook \emph{et al.} presented a method for forecasting stock market indices and their volatility.~\cite{kyoung2019performance}. Hossain \emph{et al.} proposed a hybrid deep learning model that combines Long Short-Term Memory (LSTM) and Gated Recurrent Unit (GRU).They assessed the model's prediction ability using metrics like Mean Squared Error (MSE) and Mean Absolute Percentage Error (MAP) by training and testing it on historical data from the S\&P 500 index.~\cite{hossain2018hybrid}. Ghosh \emph{et al.} provide a framework for analyzing and projecting a company's future growth using the LSTM (Long Short-Term Memory) model and the firm's net growth calculation algorithm~\cite{inproceedings}.

\section{Conclusions and Future Research}
\label{sec:conclusion}

This paper introduces a background-aware multi-source fusion financial trend forecasting mechanism (MOF), which integrates large-scale language models and time series analysis techniques to significantly enhance the accuracy and interpretability of stock price predictions. By amalgamating stock price data, policy text summaries, and stock commentary summaries, MOF constructs a multidimensional dataset that enables models to comprehensively grasp market dynamics and policy implications. The mechanism utilizes MacBERT, a pre-trained language model optimized for Chinese, to extract key information from policy texts and generate feature vectors that enrich model inputs. Employing the deep learning architecture of LSTM and its variants, MOF efficiently processes time series data and learns long-term dependencies to forecast future stock trends.

The optimization strategy, featuring the Adam optimizer and binary cross-entropy loss function, coupled with hyperparameter tuning, ensures efficient and robust model training while mitigating overfitting risks. Moreover, enhancements in model interpretability enable stakeholders to understand the rationale behind the model's predictions clearly. The ablation study further underscores the significance of the "Prior Effect," which considers a stock's rise or fall over the previous N-1 days, in enhancing forecast accuracy. Experimental results demonstrate that the MOF mechanism performs well across various evaluation metrics and introduces significant innovations, providing a robust new tool for stock market prediction and investment decision-making.

\section*{Acknowledgment}
This study was supported by the National Natural Science Foundation of China (62002067) and the Guangzhou Youth Talent of Science (QT20220101174).

\bibliographystyle{IEEEtran}
\bibliography{IEEE}

\end{document}